\documentclass[twocolumn,trackchanges]{aastex701}
\usepackage{xspace}
\usepackage{amsmath}

\newcommand{\ixpe}{\textit{IXPE}\xspace}

\begin{document}

\title{Exploring MAXI J1744$-$294: IXPE insights into a Newly Discovered X-ray Transient}

\author[orcid=0009-0001-4644-194X]{Lorenzo Marra}
\affiliation{INAF Istituto di Astrofisica e Planetologia Spaziali, Via del Fosso del Cavaliere 100, 00133 Roma, Italy}
\email[show]{lorenzo.marra@inaf.it}  

\author[0000-0001-7374-843X]{Romana Mikušincová}
\affiliation{INAF Istituto di Astrofisica e Planetologia Spaziali, Via del Fosso del Cavaliere 100, 00133 Roma, Italy}
\email{romana.mikusincova@inaf.it}

\author[0000-0002-1481-1870]{Federico M. Vincentelli}
\affiliation{INAF Istituto di Astrofisica e Planetologia Spaziali, Via del Fosso del Cavaliere 100, 00133 Roma, Italy}
\email{email}

\author[0000-0002-6384-3027]{Fiamma Capitanio}
\affiliation{INAF Istituto di Astrofisica e Planetologia Spaziali, Via del Fosso del Cavaliere 100, 00133 Roma, Italy}
\email{fiamma.capitanio@inaf.it}

\author[0000-0002-1793-1050]{Melania Del Santo}
\affiliation{INAF, Istituto di Astrofisica Spaziale e Fisica Cosmica, Via U. La Malfa 153, I-90146 Palermo, Italy}
\email{melania.delsanto@inaf.it }

\author[0000-0003-1533-0283]{Sergio Fabiani}
\affiliation{INAF Istituto di Astrofisica e Planetologia Spaziali, Via del Fosso del Cavaliere 100, 00133 Roma, Italy}
\email{sergio.fabiani@inaf.it}

\author[0000-0002-6126-7409]{Shifra Mandel}
\altaffiliation{National Science Foundation Fellow}
\affiliation{Columbia Astrophysics Laboratory, Columbia University, New York, NY 10027, USA}
\email{ss5018@columbia.edu}

\author[0000-0003-3331-3794]{Fabio Muleri}
\affiliation{INAF Istituto di Astrofisica e Planetologia Spaziali, Via del Fosso del Cavaliere 100, 00133 Roma, Italy}
\email{fabio.muleri@inaf.it}

\author[0009-0003-8610-853X]{Maxime Parra}
\affiliation{Department of Physics, Ehime University, 2-5, Bunkyocho, Matsuyama, Ehime 790-8577, Japan}
\email{maxime.parrastro@gmail.com}

\author[0000-0002-7781-4104]{Paolo Soffitta}
\affiliation{INAF Istituto di Astrofisica e Planetologia Spaziali, Via del Fosso del Cavaliere 100, 00133 Roma, Italy}
\email{paolo.soffitta@inaf.it}

\author[0009-0007-0537-9805]{Antonella Tarana}
\affiliation{INAF Istituto di Astrofisica e Planetologia Spaziali, Via del Fosso del Cavaliere 100, 00133 Roma, Italy}
\email{antonella.tarana@inaf.it}

\author[0000-0003-1285-4057]{M. Cristina Baglio}
\affiliation{INAF--Osservatorio Astronomico di Brera, Via Bianchi 46, I-23807 Merate (LC), Italy}
\email{maria.baglio@inaf.it}

\author[0000-0002-4622-4240]{Stefano Bianchi}
\affiliation{Dipartimento di Matematica e Fisica, Universit\`a degli Studi Roma Tre, Via della Vasca Navale 84, 00146, Roma, Italy}
\email{stefano.bianchi@uniroma3.it}

\author[0000-0003-4925-8523]{Enrico Costa}
\affiliation{INAF Istituto di Astrofisica e Planetologia Spaziali, Via del Fosso del Cavaliere 100, 00133 Roma, Italy}
\email{enrico.costa@inaf.it}

\author[0000-0002-5042-1036]{Antonino D'Aì}
\affiliation{INAF, Istituto di Astrofisica Spaziale e Fisica Cosmica, Via U. La Malfa 153, I-90146 Palermo, Italy}
\email{antonino.dai@inaf.it}

\author[0000-0003-2743-6632]{Barbara De Marco}
\affiliation{Departament de Fis\'{i}ca, EEBE, Universitat Polit\`ecnica de Catalunya, Av. Eduard Maristany 16, S-08019 Barcelona, Spain}
\email{barbara.de.marco@upc.edu}

\author[0000-0003-0079-1239]{Michal Dov\v{c}iak}
\affiliation{Astronomical Institute of the Czech Academy of Sciences, Bo\v{c}n\'{i} II 1401/1, 14100 Praha 4, Czech Republic}
\email{michal.dovciak@asu.cas.cz}

\author[0000-0002-9719-8740]{Vittoria Elvezia Gianolli}
\affiliation{Department of Physics and Astronomy, Clemson University, Clemson, SC, 29634, USA}
\email{vgianol@clemson.edu}

\author[0000-0002-0642-1135]{Andrea Gnarini}
\affiliation{Dipartimento di Matematica e Fisica, Universit\`a degli Studi Roma Tre, Via della Vasca Navale 84, 00146, Roma, Italy}
\email{andrea.gnarini@uniroma3.it}

\author[0000-0003-0976-8932]{Maitrayee Gupta}
\affiliation{Astronomical Institute of the Czech Academy of Sciences, Bo\v{c}n\'{i} II 1401/1, 14100 Praha 4, Czech Republic}
\email{maitrayee.gupta@asu.cas.cz}

\author[0000-0002-5311-9078]{Adam Ingram}
\affiliation{School of Mathematics, Statistics, and Physics, Newcastle University, Newcastle upon Tyne NE1 7RU, UK}
\email{adam.ingram@newcastle.ac.uk}

\author[0000-0003-4216-7936]{Guglielmo Mastroserio}
\affiliation{Dipartimento di Fisica, Univerusit\`a Degli Studi di Milano, Via Celoria, 16, Milano, 20133, Italy}
\email{guglielmo.mastroserio@unimi.it}

\author[0000-0002-2152-0916]{Giorgio Matt}
\affiliation{Dipartimento di Matematica e Fisica, Universit\`a degli Studi Roma Tre, Via della Vasca Navale 84, 00146, Roma, Italy}
\email{giorgio.matt@uniroma3.it}

\author[0000-0002-9709-5389]{Kaya Mori}
\affiliation{Columbia Astrophysics Laboratory, Columbia University, New York, NY 10027, USA}
\email{kaya@astro.columbia.edu}

\author[0000-0001-6061-3480]{Pierre-Olivier Petrucci}
\affiliation{University Grenoble Alpes, CNRS, IPAG, 38000 Grenoble, France}
\email{pierre-olivier.petrucci@univ-grenoble-alpes.fr}

\author[0000-0001-5418-291X]{Jakub Podgorn{\'y}}
\affiliation{Astronomical Institute of the Czech Academy of Sciences, Bo\v{c}n\'{i} II 1401/1, 14100 Praha 4, Czech Republic}
\email{jakub.podgorny@asu.cas.cz}

\author[0000-0002-0983-0049]{Juri Poutanen}
\affiliation{Department of Physics and Astronomy, 20014 University of Turku, Finland}
\email{juri.poutanen@utu.filim}

\author[0000-0002-5872-6061]{James~F. Steiner}\affiliation{Center for Astrophysics $\vert$ Harvard-Smithsonian, 60 Garden Street, Cambridge, MA 02138, USA}
\email{james.steiner@cfa.harvard.edu}

\author[0000-0003-2931-0742]{Ji\v{r}\'{i} Svoboda}
\affiliation{Astronomical Institute of the Czech Academy of Sciences, Bo\v{c}n\'{i} II 1401/1, 14100 Praha 4, Czech Republic}
\email{jiri.svoboda@asu.cas.cz}

\author[0000-0002-1768-618X]{Roberto Taverna}
\affiliation{Dipartimento di Fisica e Astronomia, Universit\`{a} degli Studi di Padova, Via Marzolo 8, 35131 Padova, Italy}
\email{roberto.taverna@unipd.it }

\author[0000-0002-6562-8654]{Francesco Tombesi}
\affiliation{Department of Physics, Tor Vergata University of Rome, 00133 Rome, Italy}
\email{francesco.tombesi@roma2.infn.it}

\author[0000-0002-2381-4184]{Swati Ravi}
\affiliation{MIT Kavli Institute for Astrophysics and Space Research, 77 Massachusetts Avenue, Cambridge, MA 02139, USA}
\email{swatir@mit.edu} 

\author[0000-0002-4151-4468]{J\'er\^ome Rodriguez}
\affiliation{Universit\'e Paris-Saclay, Universit\'e Paris Cité, CEA, CNRS, AIM, F9119, Gif Sur Yvette, France}
\email{jrodriguez@cea.fr}

\author[0000-0002-7930-2276]{Thomas D. Russell}
\affiliation{INAF, Istituto di Astrofisica Spaziale e Fisica Cosmica, Via U. La Malfa 153, I-90146 Palermo, Italy}
\email{thomas.russell@inaf.it}

\author[0000-0002-5767-7253]{Alexandra Veledina}
\affiliation{Department of Physics and Astronomy, 20014 University of Turku, Finland}
\affiliation{Nordita, KTH Royal Institute of Technology and Stockholm University, Hannes Alfv\'ens v\"ag 12, SE-10691 Stockholm, Sweden}
\email{alexandra.veledina@gmail.com}

\author[0000-0002-2967-790X]{Shuo Zhang}
\affiliation{Michigan State University, Department of Physics and Astronomy, East Lansing, MI, 48824}
\email{zhan2214@msu.edu}


\begin{abstract}

We present the first \ixpe spectro-polarimetric observation of the black hole candidate MAXI~J1744$-$294, a transient X-ray source discovered during a bright 2025 outburst in the Galactic center region. During the $\sim$150 ks observation, the source was found in the soft state, and its spectrum was well described by an absorbed multicolor disk with a minor high-energy tail. 
No significant polarization was detected, and we derived a $3\sigma$ upper limit on the polarization degree of $1.3\%$ in the 2--8 keV energy band. 
This result is consistent with previous findings for soft-state black hole binaries observed at low to intermediate inclination angles. By comparing the polarization degree upper limit with theoretical predictions for standard accretion disk emission, we constrain the disk inclination to $i \lesssim 38\degr$–$71\degr$, depending on the black hole spin and the disk atmosphere albedo.

\end{abstract}



\section{Introduction} 
\setcounter{footnote}{0}

Stellar mass black holes (BHs) in X-ray binary (XRB) systems serve as natural laboratories for investigating accretion physics, relativistic jet formation, and the effects of strong gravitational fields. These systems host a BH (5--30 $M_\odot$) accreting matter from a companion star via an accretion disk. While a subset of these remain persistently active over extended periods of time, the majority exhibit transient behavior, cycling between long phases of quiescence and dramatic outbursts \citep{King+96,Tetarenko+16,Corral-Santana+16}. During these outbursts, the X-ray luminosity increases by several orders of magnitude \citep{Yu+09}, providing a unique opportunity to study accretion processes across a wide range of physical scales.

A defining characteristic of BH XRBs is the pronounced variability in their spectral and timing properties throughout an outburst. This variability has led to the classification of distinct spectral states, which are interpreted as manifestations of changes in the accretion flow geometry and radiative mechanisms \citep{Zdziarski+04,Done+07}.
In the soft state, the X-ray spectrum is dominated by a multi-color blackbody component, generally interpreted as thermal emission from a geometrically thin, optically thick accretion disk extending down to the innermost stable circular orbit \cite[ISCO,][]{Shakura+73,Novikov+73,Page+74}.
Conversely, the hard state is characterized by a cutoff power-law spectrum with an energy roll-over at $\sim 100$ keV, typically attributed to Comptonization of soft disk photons by a hot, optically thin medium \cite[the corona,][]{Done+07}. Transitions between these spectral states are primarily driven by variations in the mass accretion rate, even though a second parameter (still unknown) must be at play to explain the hysteretic behavior of the outburst \citep{Belloni+10,Dunn+10,Tetarenko+16}.
These transitions typically occur on timescales ranging from hours to days.

Advancements in the X-ray spectral and timing observations allowed to outline a characteristic “q-shaped” path in the hardness-intensity diagram (HID) that sources follow during an outburst \citep{Fender+04,Homan05}. However, the physical process responsible for this hysteresis loop, in which hard-to-soft transitions occur at higher luminosities than soft-to-hard transitions \citep{Maccarone+03_hysteresis,Meyer-Hofmeister+05}, remains a critical puzzle. Intermediate states and fast transitions \citep{Belloni+05, Delsanto09} failed transition outbursts \citep{Capitanio+09,Bassi19, Alabarta+21} and the state-dependent observation of outflows such as disk winds \citep{Ponti+2012,Parra+24} and radio jets \citep{Fender+04} further complicate the picture. 

A promising avenue to disentangle these complexities is via X-ray polarimetry, which provides direct insights into the geometry and emission mechanisms of the accretion flow. This approach has become possible with the launch of the Imaging X-ray Polarimetry Explorer (\ixpe) in 2021 \citep{Weisskopf+22}, enabling the first systematic measurements of X-ray polarization from BH-XRBs in the 2--8~keV energy band. Polarization measurements of sources in the soft state can provide an additional way to constrain the spin of a BH \citep{Stark+77,Connors+77,Connors+80,Dovciak+08,Schnittman+09,Taverna+20}, while observations in the hard state offer unprecedented insights into the geometry of the corona, e.g. in comparison to the jet and the accretion-disk orientation \citep{Poutanen1996,Schnittman+10,Krawczynski+22,Zhang+22}. 

\ixpe has observed both persistent and transient BH XRBs across various spectral states \citep[see][for a recent review]{Dovciak+24}. Observations of the persistent XRB \mbox{Cyg X-1} in the hard state \citep{Krawczynski+22_cygnus,Kravtsov+25} revealed that the polarization direction is aligned with the radio jet, suggesting that the corona is extended perpendicularly to the disk symmetry axis. Similar behavior was observed during the hard-to-soft transitions of \mbox{GX~339$-$4} \citep{Mastroserio+25} and Swift~J1727.8$–$1613 \citep{Veledina+23,Ingram+24}.
The very large polarization degree recently detected in the hard state observation of the transient source \mbox{IGR J17091–3624} \cite[$9.1\%\pm1.6\%$,][]{Ewing+25}, suggests a high inclination angle for this system. However, the unknown orientation of its radio jet currently prevents a direct comparison with the polarization vector direction.
Subsequent IXPE observations of Swift J1727.8$-$1613 during its outburst revealed a gradual decrease in the polarization degree (PD) as the source transitioned toward the soft state, a very low PD in the soft state \citep{Svoboda+24}, and a re-acquisition of PD during the reverse transition back to the hard state \citep{Podgorny+24}. These findings suggest that the corona may retain a similar geometric structure during both the bright hard-to-soft and the dim soft-to-hard transitions. 

For some sources observed in soft state, only an upper limit on the PD could be established. 
Polarization measurements of \mbox{LMC X-3} \citep{Svoboda+24_LMC}, \mbox{4U~1957+11} \citep{Marra+24}, and \mbox{Cyg X-1} \citep{Steiner+24} enabled the first BH spin estimates through polarimetric analysis, which yielded results consistent with those obtained by other spectroscopic methods. 

A very large PD with a direction perpendicular to the radio jet was observed in \mbox{Cyg X-3}, varying from $\sim$10\% to $\sim$20\% across different spectral states \citep{Veledina+24,Veledina+24_Soft}, suggesting that the source is surrounded by optically thick material -- likely a dense 
wind from the accretion disk -- and allowing constraints to be placed on the geometry of the obscuring material. 
Finally, an unanticipated result came from the soft state observation of the transient source 4U~1630$-$47 \citep{Ratheesh+24}, followed by its observation in the steep power-law state \citep{RodriguezCavero+23}. 
While radio emission has recently been detected from this source \citep{Mariani+25}, it remains unresolved, so the orientation of the system is still unknown. The very high polarization level measured in the soft state ($\approx 8\%$), increasing with energy, is difficult to reconcile with the standard accretion disk model and underscores the need for further theoretical work to understand the physical processes in this source.

Here we report on an \ixpe observation of the newly discovered BH candidate \mbox{MAXI~J1744$-$294}.

\subsection{Discovery and Outburst}

On 2025 January  2 (MJD 60677), the MAXI/GSC all-sky monitor \citep{Matsuoka+09} reported a bright hard X-ray transient near the Galactic center region, initially detected at 10–20~keV with an average flux of $\sim$100~mCrab and later rising to $\sim$250~mCrab over January 13–15. 
The source was provisionally named {MAXI~J1744$-$294}. Initial analyses placed the transient position within an error ellipse overlapping Sgr~A$^*$, leading to a tentative association with the known neutron star XRB {KS~1741$-$293}. However, this association was subsequently ruled out due to the source’s unusually high flux and a hard spectral shape \citep{Atel1_Kudo,ATel2_Nakajima}.

To localize and characterize this new source, a coordinated X-ray follow-up campaign was initiated. The \textit{NinjaSat} CubeSat observed the region on 2025 January 16--23, finding a 2--10~keV flux of 22~mCrab with a steep power-law spectrum (photon index $\Gamma \approx 3.1$) and high absorption \citep[$N_{\rm H} \approx 7^{+5}_{-2} \times 10^{23}~\mathrm{cm^{-2}}$;][]{ATel3_Watanabe}.
The \textit{Swift}/XRT pointing on February 1 revealed two sources near Sgr~A$^*$: a fainter one consistent within 10\arcsec\ of the known neutron star XRB {AX~J1745.6$-$2901} (with a count rate of $\sim$0.2~count\,s$^{-1}$), and a brighter, piled-up source (count rate $\sim$6~count\,s$^{-1}$) centered 25\arcsec\ southeast of Sgr~A$^*$. Spectral analysis of the latter suggested it had transitioned to the soft state \citep[characterized by the disk blackbody temperature of $kT \approx 0.8$~keV;][]{ATel4_Heinke}.

\textit{NuSTAR}’s 22.7~ks Target of Opportunity (ToO) observation on 2025 February 6 confirmed the counterpart at a position within $\sim$20\arcsec\ of Sgr~A$^*$. The spectrum was well described by an absorbed disk blackbody plus power-law model, along with a Gaussian  component to fit a broad emission line centered at 6.63~keV ($\sigma \approx 0.2$~keV), evident in the  6--7~keV range \citep{ATel5_Mandel}.
Further soft X-ray spectroscopy and timing studies were carried out with \textit{NICER} on 2025 February 11--12. Although the observation may include some contribution from {AX~J1745.6$-$2901}, due to \textit{NICER}’s lack of imaging capability, the measured spectral parameters are consistent (within uncertainties) with those obtained by \textit{NuSTAR} \citep{ATel6_Jaisawal}.
Simultaneously, radio observations with {MeerKAT} on 2025 January 25 detected a new point source 18\arcsec\ from Sgr~A$^*$ and 23\arcsec\ from the \textit{Swift} position, confirming the first radio counterpart of {MAXI~J1744$-$294} with a peak brightness of $\sim$0.3~Jy \citep{ATel7_Grollimund}.

In March 2025, high-resolution spectroscopy with the \textit{XRISM}/Resolve instrument \citep[70~ks exposure,][]{ATel8_Mandel}, and X-ray observations with the Wide-field X-ray Telescope (WXT) and Follow-up X-ray Telescope (FXT) on board on the \textit{Einstein Probe} \citep{ATel9_Wang}, further constrained the spectral shape of this transient, estimating an absorbed flux of $\sim 8.5 \times 10^{-10}~\mathrm{erg~s^{-1}~cm^{-2}}$ in the 2–10~keV band. A \textit{Chandra}/ACIS-S ToO on March 9 achieved sub-arcsecond localization at (RA, Dec) = (17:45:40.476, $-$29:00:46.10) $\pm$ 1.2\arcsec\, solidifying the transient’s unique identification and confirming its offset by $<$20\arcsec\ from Sgr~A$^*$ \citep{ATel10_Mandel}.

The emerging picture for MAXI~J1744$-$294 is that of a bright, heavily absorbed X‑ray transient, likely a BH low‑mass XRB, undergoing a soft‑state outburst in the crowded environment of the Galactic center. This region is known to host a large population of X-ray transients, as revealed through deep \textit{Chandra} and \textit{Swift}/XRT monitoring over the past two decades \citep{Muno+05,Degenaar+15,Hailey+18}. Within approximately 100 pc of SgrA$^*$, an estimated 10--15\% of all Galactic low-mass XRBs are found, either in quiescence or in outburst, alongside a number of candidate high-mass XRBs \citep{Mori+21,Fortin+24,Mandel+25}. Thus, MAXI~J1744$-$294 joins a growing census of GC transients, providing a new observational window onto both the compact‑object density cusp around Sgr~A$^*$ and the physical mechanisms that trigger XRB outbursts.

In this work, we report on the \ixpe ToO observation of MAXI~J1744$-$294, performed in April 2025 while the source was in the soft state, along with simultaneous coverage by the \textit{Swift}/XRT and {NICER} instruments.

\section{Observation and Data reduction}

The Imaging X-ray Polarimetry Explorer (\ixpe) is the first space mission dedicated to investigating linear polarization in the 2--8 keV energy band \citep{Weisskopf+22}. It carries three co-aligned X-ray telescopes, each integrating a mirror module assembly \citep{Ramsey+22} with a gas pixel detector sensitive to polarization \citep{Baldini+21,Soffitta+21}. This design enables high-sensitivity measurements of both point-like and extended sources. The observatory achieves an angular resolution of approximately $30\arcsec$ (half-power diameter, averaged over the three detector units, DUs); the overlap of the fields of view of the three DUs is circular with a diameter of $9\arcmin$. The energy resolution is better than $20\%$ at 6 keV.

\ixpe performed a ToO observation of MAXI~J1744$-$294 from 2025 April 5 at 07:36:34 UTC to April 8 at 02:09:33 UTC (\textsc{obsid}: 04250301), for a total livetime of approximately 150~ks. We downloaded the publicly available Level 1 and Level 2 data from the \ixpe archive\footnote{\url{https://heasarc.gsfc.nasa.gov/docs/ixpe/archive/}} and analyzed them using both the \textsc{ixpeobssim} software \citep[version 31.0.3;][]{Baldini+22} and the \textsc{heasoft} package \textsc{xspec} \citep[version 12.15;][]{Arnaud+96}.
The source region was defined using the \textsc{SAOImageDS9}\footnote{\url{https://sites.google.com/cfa.harvard.edu/saoimageds9}} software. Events were extracted from a circular region with a radius of $60\arcsec$, centered on the centroid of the source counts for each \ixpe detector. Due to the modest count rate of the source during the observation ($\approx 1.14 \ \mathrm{count\,s^{-1}\, arcmin^{-2}}$ in the selected region), we applied the background rejection procedure described in \citet{DiMarco+23}.
Given the source position near the Galactic center and its proximity to AX~J1745.6$-$2901, we explored several possible background extraction regions, as detailed in Section \ref{sec:POL_measure}. Notably, AX~J1745.6$-$2901, located at $\sim1\farcm3$ from MAXI~J1744$-$294, was serendipitously included in the \ixpe field of view. The analysis of its polarization properties will be presented in a forthcoming publication (Mikušincová et al., in prep.).
The weighted Stokes $I$, $Q$, and $U$ spectra were extracted using the {\tt xselect} tool from the \textsc{heasoft} package (version 6.35.1) with the command {\tt extract spect stokes=NEFF}. Response files were generated using the {\tt ixpecalcarf} tool on the event and attitude files of each DU, applying the latest available \ixpe calibration (v13).

The \textit{Swift}/X-Ray Telescope (XRT) \citep{Burrows+05} observed MAXI~J1744$-$294 in Window Timing (WT) mode on 2025 April 5 from 02:50:00 UTC to 03:04:00 UTC (ObsID 00033979025). We reprocessed the XRT data using the task \texttt{xrtpipeline} included in the \textsc{heasoft} v.6.34 software package  and the latest XRT Calibration Database (CALDB v. 20240522). We verified that the observation is not affected by the pile-up.
We selected a circular region with a radius of 8 pixels (1 pixel = 2.36\arcsec), corresponding to approximately 50\% of the point spread function (PSF), centered on the source position to extract the spectra (and response files) using the task \texttt{xrtproducts}. We adopted this unconventional radius (8 pixels instead of the standard 20) to reduce contamination from AX~J1745.6$–$2901.

To characterize the temporal behavior of the source, quasi-simultaneous observations in the 0.5--10~keV band were also obtained with {NICER} on board of the International Space Station. The closest observation during night-time was performed on 2025 April 3 (\textsc{obsid}: 8205140121) for a total of 300~s. Clean events were extracted using the \textsc{nicerl2} software (v.~1.36) distributed by HEASARC, with the default parameters. By checking the 10-12 keV lightcurve, we noticed that the first $\approx$50 s were affected by high background, and thus were excluded from the analysis. Timing analysis was then performed using a rebinned light curve with a time resolution of 1~ms. Due to {NICER}’s lack of imaging capability, however, its data were excluded from the spectral analysis described in Section~\ref{sec:Spectral_fit}, as it was not possible to reliably disentangle the contribution from AX~J1745.6$-$2901.

\section{Polarization measurement}\label{sec:POL_measure}

\begin{figure}
\centering
\includegraphics[width=\columnwidth]{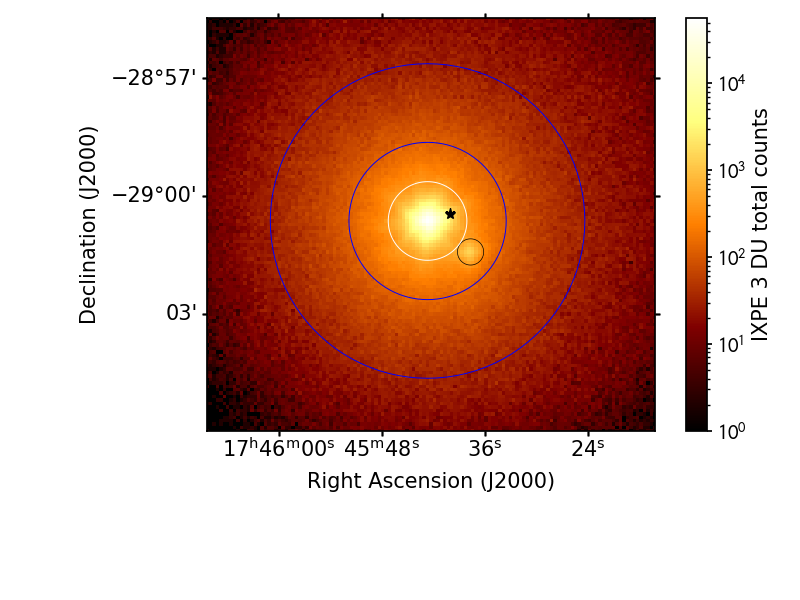}
\vspace{-10mm}
\caption{Summed count map of the three \ixpe detector units. The color scale is logarithmic to enhance the visibility of both sources and the much fainter background. The white circle indicates the source region used for extraction, while the black circle marks the position of AX~J1745.6$-$2901. The blue annulus shows one of the background regions described in Section~\ref{sec:POL_measure}; the alternative configuration uses an annulus extending from the edge of the source region to the same outer radius. The black star indicates the position of Sgr~A$^*$.}
\label{fig:IxpeCountMap}
\end{figure}

The source and background region used in the extraction of the \ixpe data are presented in Figure~\ref{fig:IxpeCountMap}, superimposed to the summed count map of the three \ixpe DUs.
We initially used the {\tt xpselect} tool from the \textsc{ixpeobssim} software to filter the counts of MAXI~J1744$-$294  region, and then generated the unweighted polarization cubes in the 2--8~keV band employing the {\tt PCUBE} algorithm of the {\tt xpbin} tool. This analysis revealed a PD lying below the minimum detectable polarization value at the 99$\%$ confidence level \citep[MDP$_{99}$,][]{Weisskopft+10}, estimated at $1.2\%$ for the observation; at 3$\sigma$ confidence level only an upper limit on the $\mathrm{PD}<1.9\%$ can be established; consequently, the polarization angle (PA) remains unconstrained. 

Subsequently, we investigated whether the polarization signal detected for MAXI~J1744$-$294 could be contaminated by diffuse emission from the Galactic center or by emission from the nearby neutron star transient AX~J1745.6$–$2901. To assess this, we extracted background counts from an annular region centered on the source position. Two configurations were tested: one excluding AX~J1745.6$-$2901 by selecting inner and outer radii of $120\arcsec$ and $240\arcsec$, respectively, and one including it, with inner and outer radii of $60\arcsec$ and $240\arcsec$. In both cases, the background-subtracted \texttt{PCUBE} analysis yielded results consistent with those obtained from the source region, with the polarization signal remaining below MDP$_{99}$. The estimated upper limits on the PD at the 3$\sigma$ confidence level are $1.9\%$ and $2.0\%$ for the two configurations, respectively. Energy-resolved polarization analyses similarly showed no significant signal above MDP$_{99}$ in any tested interval. 
To further investigate potential contamination from the Galactic center diffuse X-ray emission, we repeated the analysis using a smaller source region with a $30\arcsec$ radius. The 3$\sigma$ PD upper limits remained consistent at $2.0\%$ whether considering only the source region or subtracting either of the two background contributions. Given this consistency across source and background selections, we ultimately adopted the larger source region with a $60\arcsec$ radius and the background region excluding AXJ1745.6$-$2901 (represented by the white circle and blue annulus in Figure \ref{fig:IxpeCountMap}) for the subsequent timing and spectral analyses. We note that a more stringent $3 \sigma$ upper limit on the PD can be obtained through weighted spectro-polarimetric fitting, as discussed in Section \ref{sec:Pol_xspec}.

\begin{figure}
\centering
\includegraphics[width=\columnwidth]{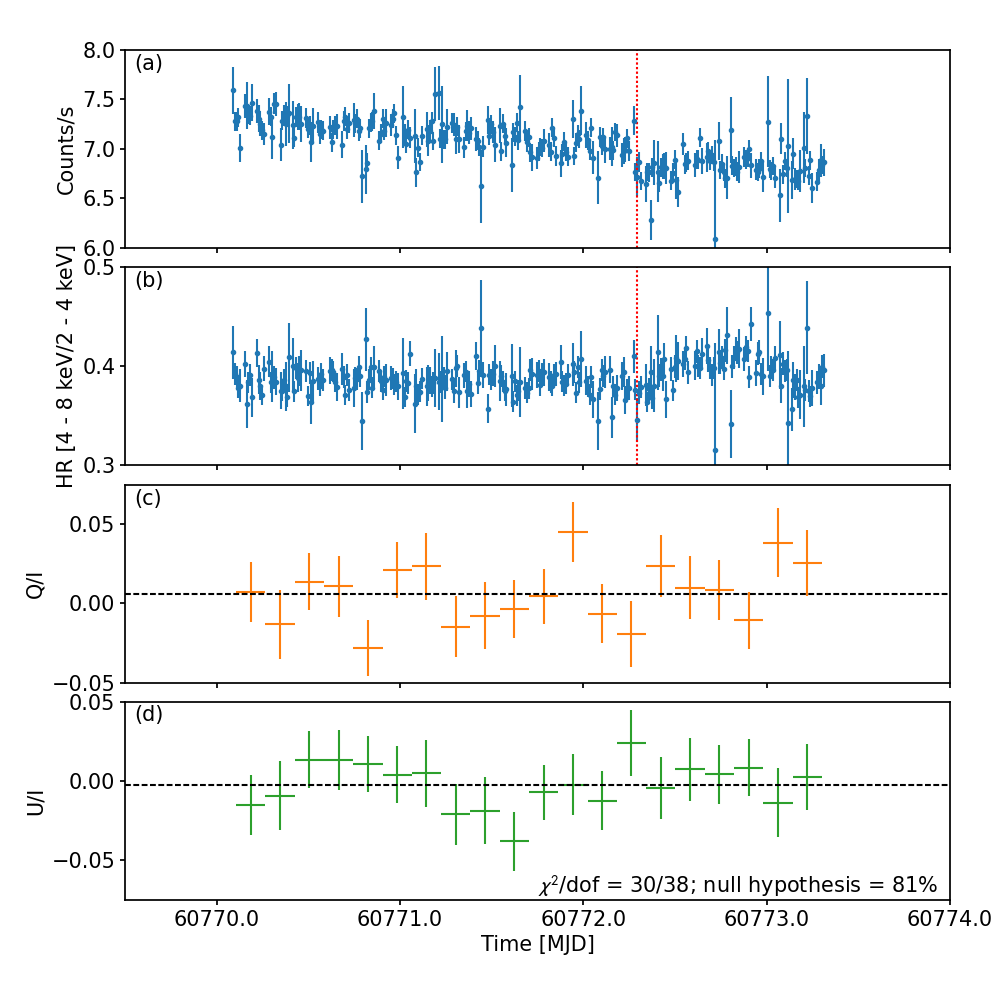}
\caption{Light curves and time-dependent polarization properties of MAXI~J1744$-$294 from the \ixpe observation.
(a) \ixpe light curve in the 2--8~keV energy range with 800~s time bins.
(b) Time evolution of the hardness ratio, defined as the ratio of \ixpe count rates in the 4--8 keV and 2--4 keV bands, using the same 800~s time bins. The vertical dotted red line in panels (a) and (b) highlights the count-rate drop described in Section~\ref{sec:timing}.
(c, d) Normalized Stokes $Q$ and $U$ parameters, measured by \ixpe as a function of time. Dashed lines show the best-fit constant model; the corresponding $\chi^2$, numbers of degrees of freedom, and null hypothesis probabilities are reported in panel (d).}
\label{fig:IXPE_lc}
\end{figure}

Although no significant time-averaged polarization is observed, a time-dependent signal could still be present in the \ixpe observation. To investigate this, we computed the normalized Stokes parameters $Q/I$ and $U/I$ in 20 time bins of approximately 7.5 ks, as shown in the two bottom panels of Figure~\ref{fig:IXPE_lc}. These quantities are expected to behave as independent normal variables \citep{Kislat+15}. We fit their temporal evolution with constant models; the resulting null hypothesis probabilities are $\sim81\%$ for both $Q/I$ and $U/I$, indicating that the observed fluctuations are statistically consistent with random noise. We therefore conclude that no significant time variability in polarization is detected.

\section{Spectro-polarimetric analysis}\label{sec:Spectral_fit}

\subsection{Temporal behaviour}\label{sec:timing}

During its outburst, MAXI~J1744$-$294 was reported to be in the soft spectral state by several observatories beginning in early February 2025 \citep{ATel4_Heinke,ATel5_Mandel,ATel9_Wang}. By the time of the \ixpe observation in early April 2025, the source flux was gradually declining. 
The 2--8 keV count rate, shown in the top panel of Figure~\ref{fig:IXPE_lc} with 800~s time bins, decreased by approximately $10\%$ over the duration of the observation. The hardness ratio, defined as the ratio of the 4--8 keV count rate to the 2--4 keV count rate and shown in  Figure~\ref{fig:IXPE_lc}b, remained stable, suggesting that no state transition occurred during the \ixpe pointing.
A sharp $\sim5\%$ drop in the count rate was detected on 2025 April 7 at approximately 06:57 UTC (MJD 60772.29), highlighted in Figure~\ref{fig:IXPE_lc}a by a vertical red dotted line. Coinciding with this event, a slight hardening in the spectrum is hinted at by the hardness ratio; however, given \ixpe's narrow energy band, no firm conclusions can be drawn. To assess any possible polarization associated with this variability, we divided the observation into two segments, before and after the drop, but in both intervals no polarization signal was detected.

To analyze the possible variability of the source, we used the NICER light curve with 1~ms time resolution in the 0.5--10~keV band. Although the short exposure  allowed us to appreciate any long-term trend, the light curve did not show any sign of stochastic variability. This was also clear from the Fourier analysis. The power density spectrum, computed using the numerical recipe described in \citet{uttley2014} with 8192 bin per segments (i.e. minimum frequency $\approx$ 0.122 Hz) showed a flat behavior, consistent with only white noise. After subtracting the Poisson noise contribution, we estimate an rms between 0.122 and 500~Hz of $\approx3\%$.

\subsection{Spectral fit}

\begin{figure}
\centering
\includegraphics[width=\columnwidth]{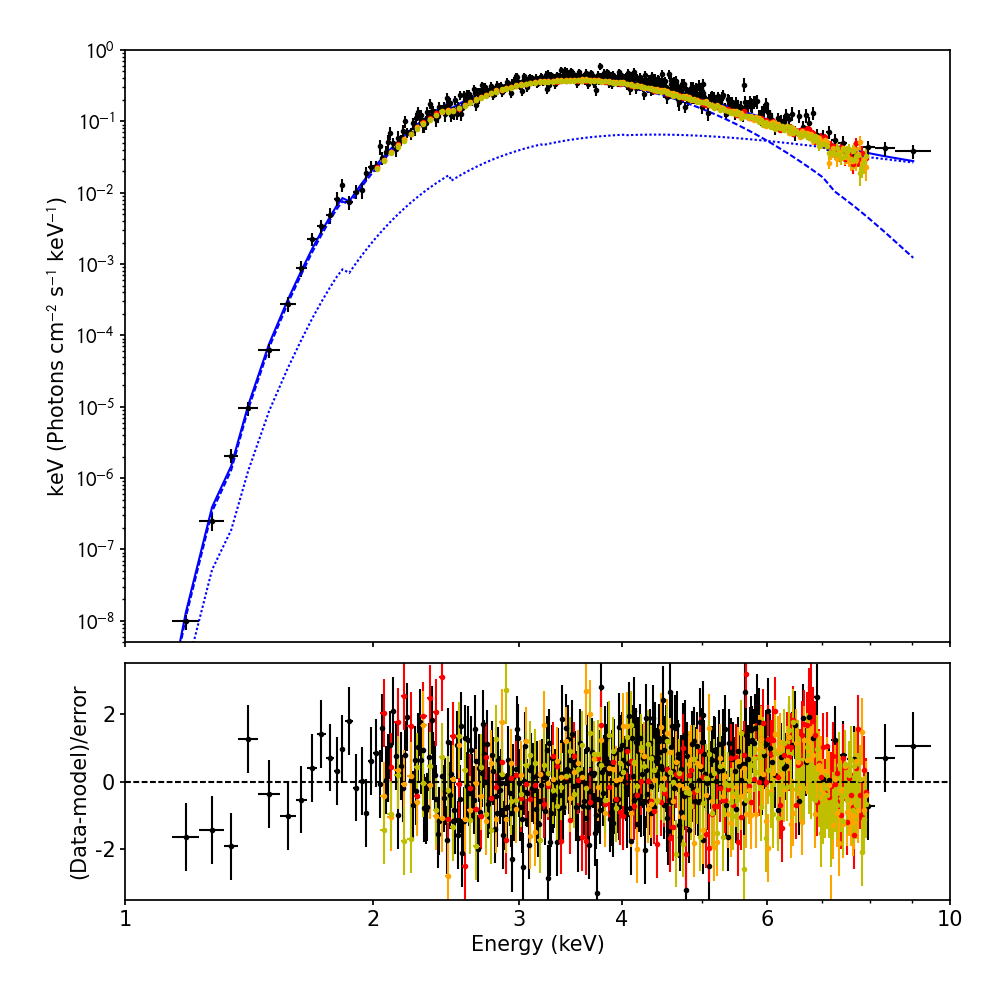}
\caption{X-ray spectra of MAXI~J1744$-$294 as observed by \ixpe and \textit{Swift}/XRT. Top panel: unfolded spectra (i.e. the flux $F(E)$) for the best-fit model. Swift/XRT data are shown in black and the three IXPE DUs in red, orange, and yellow, respectively. The blue curves represent the best-fit spectral model: the solid line indicates the total emission, while the dashed and dotted lines correspond to the disk and Comptonized components, respectively. Bottom panel: data minus model residuals in units of $\sigma$.}
\label{fig:IXPE_spectra}
\end{figure}

Due to the overall stability of the hardness ratio throughout the observation, we used time-averaged \ixpe data for the subsequent spectral analysis in \textsc{xspec}. We performed a joint fit of the three \ixpe Stokes $I$ spectra in the 2--8~keV band together with the \textit{Swift}/XRT spectrum in the 1--10~keV range. The fit statistics are presented in Table~\ref{tab:spectralfit}, while the unfolded spectra along with the data-model residuals are shown in Figure \ref{fig:IXPE_spectra}.

The spectrum of MAXI~J1744$-$294 is well described by an absorbed disk blackbody with a minor contribution from a Comptonized component, consistent with the source being in the soft state. We  fit the data employing a model consisting of a \texttt{diskbb} component \citep{Mitsuda+84}, representing a multi-temperature black-body accretion disk, and a \texttt{powerlaw} component to describe the high-energy tail. Interstellar absorption was modeled using the {\tt tbabs} model with the relative abundances from \citet{Wilms+00}. We additionally included a \texttt{constant} component in each fit to account for cross-calibration between \textit{Swift}/XRT and the \ixpe DUs. This constant was  fixed at 1 for \textit{Swift}, and left free to vary independently for the three \ixpe DUs. 
The fit results in a $\chi^2$/dof of $871/808$. The inner disk temperature was $0.64 \pm 0.01$~keV, in agreement with the previous estimates by \citet{ATel5_Mandel,ATel8_Mandel}. 
For the interstellar absorption we estimated a hydrogen column density of $N_{\rm H} = (19.2^{+0.6}_{-0.3}) \times 10^{22} \ \mathrm{cm^{-2}}$. This value is notably larger than initial estimates for the source \cite[$N_\mathrm{H} \sim 11$–$13 \times 10^{22} \ \mathrm{cm^{-2}}$,][]{ATel5_Mandel,ATel8_Mandel,ATel9_Wang}, but is consistent with more recent results from \textit{XRISM} and \textit{XMM-Newton} observations (Parra et al., in prep; Mandel et al., in prep). 
Due to the limited high-energy coverage of both instruments employed in our analysis, we find that both the photon index and the normalization of the \texttt{powerlaw} are largely unconstrained.

From this analysis, we estimate a 2--8~keV absorbed and unabsorbed flux of $(1.64 \pm 0.02) \times 10^{-9} \ \mathrm{erg \ s^{-1} \ cm^{-2}}$ and $(1.04 \pm 0.01) \times 10^{-8} \ \mathrm{erg \ s^{-1} \ cm^{-2}}$, respectively.\footnote{Fluxes have been estimated with the \texttt{cflux} model in \textsc{xspec}} Assuming a source distance of 8~kpc, consistent with its location near the Galactic center, this corresponds to an X-ray luminosity of $L_{\rm X}\sim 4 \times 10^{38}\,\mathrm{erg\,s^{-1}}$. In this empirical decomposition, the disk component dominates the spectrum below 5~keV, contributing approximately $90\%$ of the total flux. At higher energies, the power-law component becomes more prominent, with the disk contributing only about $50\%$ of the total flux between 5 and 8 keV.

\begin{table*}
\begin{center}
\begin{tabular}{p{2cm}p{3cm}p{4.5cm}p{3cm}p{3cm}}
\hline
\hline
Component & Parameter (unit) & Description & Value Fit 1 & Value Fit 2 \\  
\hline
\hline
{\tt tbabs} & $N_\textrm{H}$ ($10^{22}$\,cm$^{-2}$) & Hydrogen column density & $19.2\substack{+0.6 \\ -0.3}$ & $19.3\substack{+0.4\\ -0.3}$ \\

\hline
{\tt diskbb} & $k T_{\rm bb}$ (keV) & Inner disk radius temperature & $0.64 \pm 0.01$ & - \\
             & norm ($10^4$)    & Normalization                 & $1.01\substack{+0.12 \\ -0.10}$ & - \\

\hline
{\tt kerrbb} & $\eta$ & Inner-torque modification & - & $0$ (frozen) \\
             & $a/M$ & BH spin & - & $0.49\substack{+0.07 \\ -0.22}$ \\
             & $i$ & Inclination (deg) & - & $<21.2$ \\
             & $M_{\rm BH}$ & BH mass ($M_\odot$) & - & $10$ (frozen) \\
             & $\dot{M}$ & Mass accretion rate ($10^{18}$ g s$^{-1}$) & - & $4.0\substack{+0.8 \\ -0.1}$ \\
             & $D$ & Distance (kpc) & - & $8$ (frozen) \\
             & $hd$ & hardening factor & - & $1.7$ (frozen) \\
             
\hline
{\tt powerlaw} & $\Gamma$ & Photon index & $2.9\substack{+0.8\\-0.7}$ & $3.1\substack{+0.3\\-0.4}$ \\
               & norm     & Normalization & $2.2\substack{+9.2 \\ -1.6}$ & $3.4\substack{+7.6 \\ -2.6}$ \\

\hline
{\tt constant} & C$_{\mathrm{DU1}}$ & Normalization & $0.89 \pm 0.01$ & $0.88 \pm 0.01$ \\
               & C$_{\mathrm{DU2}}$ & Normalization & $0.90 \pm 0.01$ & $0.90 \pm 0.01$\\
               & C$_{\mathrm{DU3}}$ & Normalization & $0.87 \pm 0.01$ & $0.87 \pm 0.01$ \\

\hline
              & $\chi^2$/dof &  & $871/808$ & $869/807$ \\

\hline
\end{tabular}
\caption{Best-fit parameters (with uncertainties at $90\%$ confidence level) of the joint {\it IXPE} and {\em Swift}/XRT spectral modeling using the combined model described in Section~\ref{sec:Spectral_fit}.}
\label{tab:spectralfit}
\end{center}
\end{table*}

We also performed a spectral fit using the relativistic disk model \texttt{kerrbb} \citep{Li+05} in place of the simpler \texttt{diskbb}, to obtain a more physically motivated description of the accretion disk emission. Because continuum-fitting methods suffer from strong degeneracies between spin, inclination, mass, and distance \citep{Remillard+06}, we fixed the source distance to $D = 8$~kpc and BH mass to $M = 10M_\odot$. We also adopted a standard disk hardening factor of $1.7$ \citep{Shimura+95}.
This model provides a good fit to the data, with $\chi^2/\mathrm{dof} = 869/807$. The fit yields a BH spin parameter of $a \approx 0.5$, while only resulting in an upper limit estimate for the inclination of $i \lesssim 21\degr$. However, we emphasize that the spin and the inclination are strongly degenerate and the associated systematic uncertainties far exceed the statistical errors listed in Table~\ref{tab:spectralfit}. These estimates are also sensitive to the assumed mass and distance. Tighter constraints on the physical parameters will require independent measurements of the source mass, distance, or inclination, or a more comprehensive analysis of the source broadband X-ray emission during the outburst (Mandel et al., in prep).

\subsection{Polarimetric fit}\label{sec:Pol_xspec}

We then incorporated the polarimetric information provided by \ixpe into our fitting procedure. Keeping all spectral parameters fixed at the values listed in Table~\ref{tab:spectralfit}, we added the \ixpe $Q$ and $U$ spectra to our analysis, applying the same cross-calibration factors used for the $I$ spectrum. We assigned a constant PD and PA to the entire model using a \texttt{polconst} component, with only the PD and PA left as free parameters during the fit. We performed this analysis over the full \ixpe energy range (2--8 keV), as well as in two smaller intervals: 2--5 keV and 5--8 keV. This choice is justified by the different flux contribution of the two spectral components in these two energy ranges, with the disk emission dominating at low energies, which is expected to be less polarized than the Comptonized radiation from the corona.

\begin{figure}
\centering
\includegraphics[width=\columnwidth]{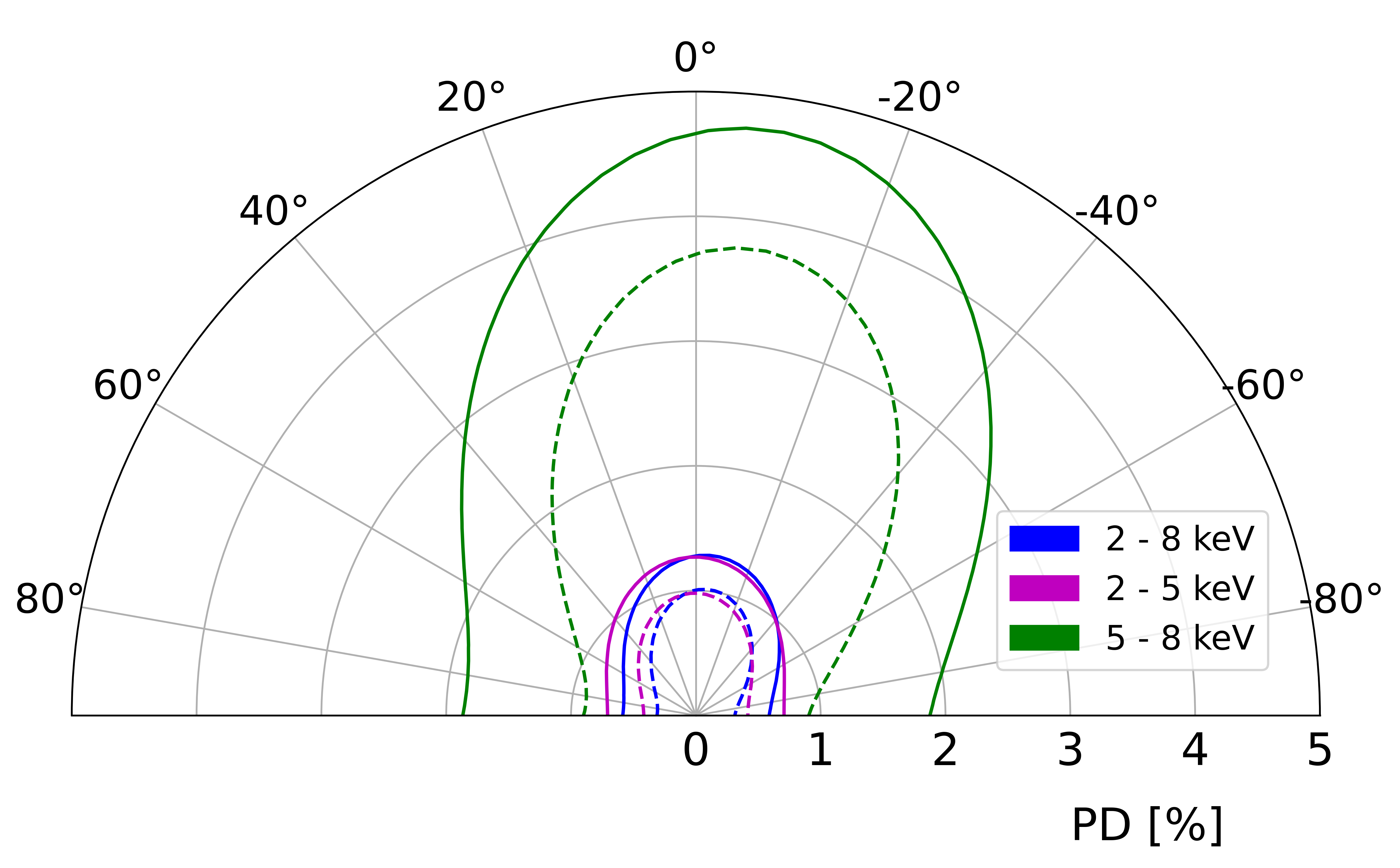}
\caption{Polar plot of the PD and PA, assuming the best-fit spectral model. Results are shown for the full \ixpe energy band (2--8 keV) as well as for two smaller intervals: 2--5 keV and 5--8 keV. Dashed and solid lines represent the $90\%$ and $99.9\%$ confidence contours, respectively.}
\label{fig:IXPE_ContourPlot_POL}
\end{figure}

Figure~\ref{fig:IXPE_ContourPlot_POL} shows the resulting PD--PA contours, computed using the \texttt{steppar} command in \textsc{xspec} with 50 steps per parameter. Across all energy ranges, we find no significant polarization detection, consistent with the \texttt{PCUBE} results presented in Section~\ref{sec:POL_measure}. We derive upper limits on the PD of $<1.3\%$ (at $3\sigma$ confidence) for both the 2--8 keV and 2--5 keV bands, increasing to PD $<4.7\%$ in the 5--8 keV band.
Notably, the upper limit obtained for the full \ixpe band is significantly lower than that from the \texttt{PCUBE} analysis presented in Section \ref{sec:POL_measure}. This difference is due to the weighted fitting approach applied in this spectro–polarimetric analysis \citep{DIMarco+22}.

\section{Discussion}

In this study, we have presented the first X-ray spectro-polarimetric analysis of MAXI~J1744$-$294. This transient source, recently discovered near the Galactic center, has been the focus of an intensive observational campaign across multiple wavelengths \citep{Atel1_Kudo,ATel2_Nakajima,ATel3_Watanabe,ATel4_Heinke,ATel5_Mandel,ATel6_Jaisawal,ATel7_Grollimund,ATel8_Mandel,ATel9_Wang,ATel10_Mandel}. Our \ixpe observation, conducted approximately three months after the outburst discovery, adds a new piece to this effort through X-ray polarimetry. At the time of the observation, the source was in the soft state, with the emission well modeled by a heavily absorbed multicolor disk component and a weak powerlaw component extending to the hard X-rays.
The X-ray flux showed a slow 
decline over the $\sim$150 ks exposure, hinting at the source entering the decay phase of the outburst.

The X-ray emission from MAXI~J1744$-$294 was found to be unpolarized, with a $1.3\%$ $3\sigma$ upper limit on the PD in both the full 2--8 keV band and the 2--5 keV range, where the disk component dominates the flux. 

Due to the source location, this polarimetric signal could be potentially contaminated by the Galactic Center diffuse X-ray emission. \ixpe has previously observed a subset of the molecular clouds composing the Sgr~A$^*$ complex \citep{Marin+23}, revealing a polarization signal in the 4--8 keV range. This signal was attributed to radiation emitted in a past flare of Sgr~A$^*$ and reflected off the dense clouds in the region, which allowed for an estimation of the flare age at $205_{-30}^{+50}$ years. Diffuse emission in the 2--4 keV energy band is instead primarily expected to be unpolarized thermal emission. Although the diffuse contribution from the Sgr~A$^*$ complex may indeed affect our observation, the large $4.7\%$ $3\sigma$ PD upper limit we estimate between 5 and 8 keV suggests that any polarization contamination is negligible.

This polarization measurement is consistent with previous \ixpe observations of accreting BH XRBs in the soft state. In general, sources observed in this spectral state have shown very low PD, with several measurements resulting in upper limits: for instance, \mbox{LMC X-1} \cite[PD $<2.5\%$;][]{Podgorny+23}, Swift~J1727.8$-$1613 \cite[PD $<1.1\%$;][]{Svoboda+24}, and \mbox{GX~339$-$4} \cite[PD $<1.4\%$;][]{Mastroserio+25}. The case of Swift~J1727.8$-$1613 is particularly illustrative, as the source was tracked throughout its outburst by \ixpe, allowing for the first time the observation of a progressive decrease in PD as the spectrum softened \citep{Ingram+24, Svoboda+24}. Polarization detections in the soft state have so far been limited to a few sources, typically characterized by large (measured or inferred) inclination angles between the observer’s line of sight and the disk axis. Notable examples include \mbox{LMC X-3} \cite[$i \approx 69\degr$;][]{Svoboda+24_LMC}, 4U~1957+11 \cite[$i \sim 50\degr$--$75\degr$;][]{Marra+24}, and 4U~1630$-$47 \cite[$i \sim 65\degr$;][]{Ratheesh+24}. An exception to this trend is \mbox{Cyg~X-1}, for which a significant polarization signal was detected in the soft state \citep{Steiner+24} despite its relatively low binary inclination \cite[$i \approx 27\degr$;][]{Orosz+11}. However, it has been proposed that the innermost accretion disk in this system may be misaligned with the binary plane, leading to a higher effective inclination that could explain the observed polarization, particularly during the hard state \cite[][but see also \citealt{Poutanen+23,Kravtsov+25}]{Krawczynski+22_cygnus}.

Disk inclination is a leading factor influencing the PD of the thermal X-ray emission from accretion disks around stellar-mass BHs. This dependence arises from the axisymmetric geometry of the disk: for a face-on configuration ($i = 0\degr$), the symmetry leads to complete cancellation of the polarization vectors, resulting in a zero net polarization. As the viewing angle increases toward an edge-on orientation ($i = 90\degr$), this symmetry is progressively lost, leading to stronger net polarization signals. In the case of MAXI~J1744$-$294, no reliable estimate of the inclination angle is currently available, given its recent discovery. However, our measured upper limit on the PD provides a useful observational constraint. By comparing this limit with theoretical predictions, we can place an initial constraint on the range of viewing angles for the system.

In the soft state sources, the observed X-ray polarization 
presumably arises from Thomson scattering in the optically thick accretion disk atmosphere. This process can be modeled using classical results derived for pure electron-scattering, semi-infinite atmospheres \citep{Chandrasekhar+60, Sobolev+63}. Additionally, in the  BH vicinity, relativistic effects are expected to significantly alter the spectro-polarimetric properties of the emerging radiation \citep{Stark+77, Connors+77, Connors+80,Dovciak+08,Taverna+20,Loktev24}. These effects lead to a net depolarization of the signal and introduce the additional contribution from the returning radiation \citep[i.e., photons emitted by the disk that are gravitationally bent and re-interact with the disk surface before reaching the observer;][]{Schnittman+09}. Returning photons are expected to scatter off the disk surface and become polarized along the disk axis, while photons emerging from the optically thick atmosphere and directly reaching the observer are predicted to be polarized perpendicularly to the disk axis \citep{Dovciak+08, Taverna+21}. 

To model these effects, we adopt the \textsc{kynbbrr} model \citep{Dovciak+08, Taverna+20, Mikusincova+23}, a relativistic ray-tracing code that simulates the spectro-polarimetric emission from BH accretion disks. The model assumes a standard Novikov–Thorne \citep{Novikov+73} disk with self-irradiation and allows us to study the variation of the observed Stokes parameters depending on the BH spin, the system inclination, and the disk atmosphere albedo, which determines the amount of returning radiation contributing to the observed spectra. 
This model has been successfully used to model the spectro-polarimetric properties of the accretion disk emission in sources observed in soft state like \mbox{LMC X-3} \citep{Svoboda+24_LMC} and  \mbox{4U~1957+11} \citep{Marra+24}. Moreover, it was applied to analyze the soft state \ixpe observation of \mbox{4U~1630-47} \citep{Ratheesh+24}, although it could not reproduce the very large PD observed in this source.

\begin{figure}
\centering
\includegraphics[width=\columnwidth]{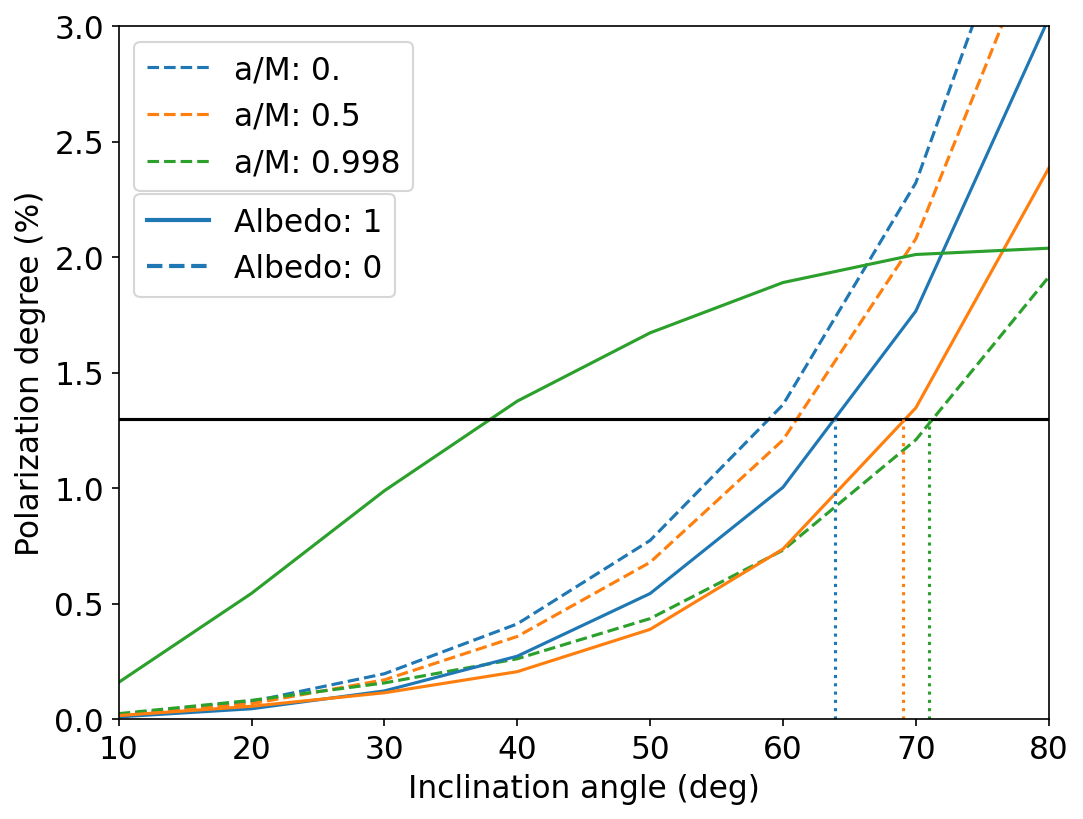}
\caption{Modeling of the PD predicted by the relativistic accretion disk model \textsc{kynbbrr} in the 2--5 keV energy range for different BH spin values: $a/M = 0$ (blue), $0.5$ (orange), and $0.998$ (green). Solid lines correspond to albedo $= 1$ and dashed lines to albedo $= 0$. The horizontal black line indicates the observed PD upper limit at the $3\sigma$ confidence level. Vertical dotted lines mark the maximum inclination allowed by the simulations for each spin value ($64\degr$, $69\degr$, and $71\degr$ for spin $a/M = 0$, $0.5$, and $0.998$, respectively), based on intersections with the $3\sigma$ line.}
\label{fig:KYN_simulations}
\end{figure}

Figure~\ref{fig:KYN_simulations} shows the dependence of the observed PD on the inclination angle, as predicted by the \textsc{kynbbrr} model for various BH spin values. The model was convolved with a \textsc{tbabs} component in \textsc{xspec} in order to take into consideration the large interstellar absorption characterizing this source. We fixed the hydrogen column density, the BH mass, its distance and its accretion rate to the values obtained in the \textsc{kerrbb} spectral fit, detailed in Table~\ref{tab:spectralfit}.
The Stokes parameters were integrated over the 2--5 keV band, where the disk emission is the main contributor to the flux. Two limiting cases for the disk surface albedo were considered: albedo $= 0$, corresponding to full absorption of returning radiation, and albedo $= 1$, where all returning radiation is reflected with no absorption. The model predictions are compared to the measured $3\sigma$ upper limit on the PD in this energy band, with the intersections representing upper limits on the inclination angle. When no returning radiation is included (albedo $= 0$), the energy-integrated PD decreases with increasing BH spin due to stronger relativistic depolarization effects. The corresponding inclination upper limits are $i < 59\degr$, $61\degr$, and $71\degr$ for spin parameters $a/M = 0$, $0.5$, and $0.998$, respectively. 

The inclusion of returning radiation contribution (albedo $=1$) results in the predicted PD decreasing when the BH spin is $0$ or $0.5$, due to the mixing of two orthogonally polarized contribution. Conversely, for a rapidly spinning black hole ($a/M = 0.998$), the returning radiation becomes more dominant, even in the 2–5 keV band, leading to an overall increase in the predicted PD. In this configuration, the inferred inclination limits are $i < 64\degr$, $69\degr$, and $38\degr$ for spin values of $0$, $0.5$, and $0.998$, respectively.
This analysis of the polarimetric data indicates that MAXI~J1744$-$294 is observed at a relatively low/intermediate inclination angle, in agreement with the fact that no eclipses, dips in the X-ray light curves, or wind signatures in the X-ray spectra have been observed to date.


\section{Conclusions}

We presented the first \ixpe spectro-polarimetric observation of the black hole candidate MAXI~J1744$-$294 during its 2025 outburst. The source was observed in the soft state, with a spectrum dominated by a multicolor disk and a weak Comptonized tail. No significant polarization signal was detected, and we derived a $3\sigma$ upper limit of $1.3\%$ on the PD in both the 2--8 keV and the 2--5 keV bands. This result is consistent with previous \ixpe measurements of low to moderate inclination BH XRBs in similar states. By comparing our PD upper limit with the theoretical prediction from relativistic disk models, we constrain the inclination of the system to $i \lesssim 38\degr$–$71\degr$, depending on the black hole spin and disk atmosphere albedo. Our findings, in line with other \ixpe observations, confirm the trend of low polarization signals in sources observed in soft state, and underscore the role of X-ray polarimetry in revealing key aspects of BH accretion geometry.

\begin{acknowledgments}
The Imaging X-ray Polarimetry Explorer (IXPE) is a joint US and Italian mission.  The US contribution is supported by the National Aeronautics and Space Administration (NASA) and led and managed by its Marshall Space Flight Center (MSFC), with industry partner Ball Aerospace (contract NNM15AA18C).  The Italian contribution is supported by the Italian Space Agency (Agenzia Spaziale Italiana, ASI) through contract ASI-OHBI-2022-13-I.0, agreements ASI-INAF-2022-19-HH.0 and ASI-INFN-2017.13-H0, and its Space Science Data Center (SSDC) with agreements ASI-INAF-2022-14-HH.0 and ASI-INFN 2021-43-HH.0, and by the Istituto Nazionale di Astrofisica (INAF) and the Istituto Nazionale di Fisica Nucleare (INFN) in Italy.  This research used data products provided by the IXPE Team (MSFC, SSDC, INAF, and INFN) and distributed with additional software tools by the High-Energy Astrophysics Science Archive Research Center (HEASARC), at NASA Goddard Space Flight Center (GSFC).

The work of L.M., S.F., F.M., P.S., G.Matt, and R.T. is partially supported by the PRIN 2022 - 2022LWPEXW - “An X-ray view of compact objects in polarized light”, CUP C53D23001180006.
M.D., M.G., J.Pod. and J.S. thank GACR project 21-06825X for the support and institutional support from RVO:67985815. 
VEG acknowledges funding under NASA contract 80NSSC24K1403. F.M.V. is supported by  the European Union’s Horizon Europe research and innovation programme through the Marie Sk\l{}odowska-Curie grant agreement No. 101149685. P.O.P. acknowledges financial support from the CNRS ``Action Th\'ematique Processus Extr\^emes et Multimessagers'' and from the CNES, the French Space Agency. A. I. acknowledges support from the Royal Society.
AV acknowledges support from the Academy of Finland grant 355672.
Nordita is supported in part by NordForsk.
MDS acknowledges ASI-INAF program I/004/11/6 (Swift). AT, FC, and SF  acknowledge financial support by the Istituto Nazionale di Astrofisica (INAF) grant 1.05.23.05.06: ``Spin and Geometry in accreting X-ray binaries: The first multi frequency spectro-polarimetric campaign''
\end{acknowledgments}

\bibliography{sample701}{}
\bibliographystyle{aasjournalv7}



\end{document}